\documentclass{aa}
\input psfig.tex
\usepackage{graphicx}
\usepackage{natbib}
\bibpunct{(}{)}{;}{a}{}{,}
\usepackage{txfonts}
\usepackage{array}
\usepackage{graphics}
\usepackage{latexsym}
\usepackage{amssymb}
\usepackage{amsmath}
\usepackage{fancyhdr}
\usepackage{float}
\usepackage{morefloats}
\usepackage{slashbox}
\usepackage{multirow}
\usepackage[toc,page]{appendix}
\bibpunct{(}{)}{;}{a}{}{,}
\include{hyphe}

\usepackage[english]{babel}

\begin{document}
\title{The ASTRODEEP Frontier Fields Catalogues: I - Multiwavelength photometry of Abell-2744 and MACS-J0416}

\author{E. ~Merlin \inst{1}
\and R. ~Amor\'{\i}n \inst{1}
\and M. ~Castellano \inst{1}
\and A. ~Fontana \inst{1}
\and F. ~Buitrago \inst{2,10,11}
\and J. ~S. ~Dunlop \inst{2}
\and D. ~Elbaz \inst{3}
\and A. ~Boucaud \inst{4,12}
\and N. ~Bourne \inst{2}
\and K. ~Boutsia \inst{1}
\and G. ~Brammer \inst{5}
\and V. ~A. ~Bruce \inst{2}
\and P. ~Capak \inst{6}
\and N. ~Cappelluti\inst{7}
\and L. ~Ciesla \inst{3}
\and A. ~Comastri \inst{7}
\and F. ~Cullen \inst{2}
\and S. ~Derriere \inst{8}
\and S. ~M. ~Faber \inst{9}
\and H.~ C. ~Ferguson \inst{5}
\and E. ~Giallongo \inst{1}
\and A. ~Grazian \inst{1}
\and J. ~Lotz \inst{5}
\and M. ~J. ~Micha{\l}owski \inst{2}
\and D. ~Paris \inst{1}
\and L. ~Pentericci \inst{1}
\and S. ~Pilo \inst{1}
\and P. ~Santini \inst{1}
\and C. ~Schreiber \inst{3,13}
\and X. ~Shu \inst{3,14}
\and T. ~Wang \inst{3,15}
}

\institute{INAF - Osservatorio Astronomico di Roma, Via Frascati 33, I - 00078 Monte Porzio Catone (RM), Italy
\email{emiliano.merlin\char64oa-roma.inaf.it}\label{inst1}
\and SUPA\thanks{Scottish Universities Physics Alliance}, Institute for Astronomy, University of Edinburgh, Royal Observatory, Edinburgh, EH9 3HJ, U.K. \label{inst2}
\and Laboratoire AIM-Paris-Saclay, CEA/DSM/Irfu - CNRS - Universit\'e Paris Diderot, CEA-Saclay, pt courrier 131, F-91191 Gif-sur-Yvette, France \label{inst3}
\and IAS - Institut d'Astrophysique Spatiale Universit\'{e} Paris Sud, B\^atiment 121, 91405 Orsay, France \label{inst4}
\and Space Telescope Science Institute, 3700 San Martin Drive, Baltimore, MD 21218, USA \label{inst5}
\and Spitzer Science Center, 314-6 Caltech, Pasadena, CA 91125, USA \label{inst6}
\and INAF - Osservatorio Astronomico di Bologna, Via Ranzani 1, I - 40127, Bologna, Italy \label{inst7}
\and Observatoire astronomique de Strasbourg, Universit\'e de Strasbourg, CNRS, UMR 7550, 11 rue de l'Universit\'e, F-67000 Strasbourg, France \label{inst8} 
\and UCO/Lick Observatory, University of California, 1156 High Street, Santa Cruz, CA 95064, USA \label{inst9}
\and Instituto de Astrof\'{\i}sica e Ci\^{e}ncias do Espa\c{c}o, Universidade de Lisboa, OAL, Tapada da Ajuda, PT1349-018 Lisbon, Portugal\label{inst10} 
\and Departamento de F\`{i}sica, Faculdade de Ci\^{e}ncias, Universidade de Lisboa, Edif\`{i}cio C8, Campo Grande, PT1749-016 Lisbon, Portugal \label{inst11}
\and Sorbonne Universit\'es, UPMC Univ Paris 6 et CNRS, UMR 7095,  Institut d’Astrophysique de Paris, 98 bis bd Arago, 75014 Paris, France \label{inst12}
\and Leiden Observatory, Leiden University, NL-2300 RA Leiden, The Netherlands \label{inst13}
\and Department of Physics, Anhui Normal University, Wuhu, Anhui, 241000, China \label{inst14}
\and School of Astronomy and Astrophysics, Nanjing University, Nanjing, 210093, China \label{inst15}
}

\abstract
{
The Frontier Fields survey is a pioneering observational program aimed at collecting photometric data, both from space (Hubble Space Telescope and Spitzer Space Telescope) and from ground-based facilities (VLT Hawk-I),  for six deep fields pointing at clusters of galaxies and six nearby deep parallel fields, in a wide range of passbands. The analysis of such data is a natural outcome of the \textsc{Astrodeep} project, an EU collaboration aimed at developing methods and tools for extragalactic photometry and create valuable public photometric catalogues.
}
{
We produce multiwavelength photometric catalogues (from $B$ to 4.5 $\mu$m) for the first two of the Frontier Fields, Abell-2744 and MACS-J0416 (plus their parallel fields).
}
{
To detect faint sources even in the central regions of the clusters, we develop a robust and repeatable procedure that uses the public codes \textsc{Galapagos} and \textsc{Galfit} to model and remove most of the light contribution from both the brightest cluster members as well as the Intra-cluster Light. We perform the detection on the processed HST \textit{H}160 image to obtain a pure \textit{H}-selected sample, that is the primary catalog that we publish. We also add a sample of sources which are undetected in the \textit{H}160 image but appear on a stacked infrared image.
Photometry on the other HST bands is obtained using \textsc{SExtractor}, again on processed images after the procedure for foreground light removal. Photometry on the Hawk-I and IRAC bands is obtained using our PSF-matching deconfusion code \textsc{t-phot}. A similar procedure, but without the need for the foreground light removal, is adopted for the Parallel fields.
}
{
The procedure of foreground light subtraction allows for the detection and the photometric measurements of $\sim 2500$ sources per field. We deliver and release complete photometric $H$-detected catalogues, with the addition of the complementary sample of infrared-detected sources. All objects have multiwavelength coverage including \textit{B} to \textit{H} HST bands, plus \textit{K}-band from Hawk-I, and 3.6 - 4.5 $\mu$m from Spitzer. Full and detailed treatment of photometric errors is included. We perform basic sanity checks on the reliability of our results.
}
{
The multiwavelength photometric catalogs are publicly available and are ready to be used for scientific purposes. Our procedures allows for the detection of outshined objects near the bright galaxies, which, coupled with the magnification effect of the clusters, can reveal extremely faint high redshift sources. Full analysis on photometric redshifts is presented in a companion Paper II.
}

\authorrunning{E. Merlin, R. Amorin, M. Castellano et al.}
\titlerunning{ASTRODEEP Frontier Fields I - Photometry}

\maketitle

\section{Introduction}\label{Introduction}

Multiwavelength photometric catalogues are a fundamental tool to investigate the properties of high-redshift galaxies, where large statistical spectroscopic studies are unfeasible and morphology is unclear. With the aid of the most powerful available telescopes, like the Hubble Space Telescope (HST), the Spitzer Space Telescope, and ground based facilities like e.g. the Large Binocular Telescope and the Very Large Telescope, it has been recently possible to build large libraries of photometric data on the faintest and most distant sources in selected regions of the sky \citep[see e.g.][]{Guo2013, Agueros2005, Obric2006, Grogin2011}.

The Frontier Fields (FF) progam \citep{Lotz2014, Koekemoer2014} offers a unique opportunity to obtain high quality data of unprecedented depth. Thanks to the natural magnification provided by the foreground galaxy clusters targeted by the observations, it is possible to observe extremely faint galaxies enhanced in luminosity by the gravitational lensing provided by the clusters' mass. The vast scale of the program (six cluster fields plus six parallel fields) also reduces cosmic variance effects in the analysis of data from previous surveys. Moreover, the program offers a crucial test for our current capabilities of data analysis, in the perspective of the new, large amount of data coming from future facilities like the James Webb Space Telescope.
Published works on FF data include, among others, \citet{McLeod2015,WangX2015,Oesch2015,Laporte2015,Atek2015}.

However, the conceptual and technical challenges in the analysis of these data are huge. Combining all the available data at different wavelengths and resolution is \textit{per se} an extremely difficult task. On top of that, the presence of bright cluster members and the Intra Cluster Light (ICL) make the whole procedure more complicated: they cover a significant fraction of the highly magnified region, outshining the faintest sources and making the whole background variable and difficult to estimate. 

A possible way to fully exploit the depth of the observations that we explore in this paper is to try to artificially ``removing'' such objects from the field of view, via analytic fitting. We present a complete multiwavelegth photometric catalogue for the first two observed FF, Abell-2744 (A2744 hereafter) and MACS-J0416 (M0416 hereafter) (two cluster fields plus two parallel fields), along with a detailed description of the methodology we have developed to obtain such photometric data. The catalogues have been obtained using techinques and software developed within the \textsc{Astrodeep} project\footnote{\textsc{astrodeep} is a coordinated and comprehensive program of i) algorithm/software development and testing; ii) data reduction/release, and iii) scientific data validation/analysis of the deepest multi-wavelength cosmic surveys. For more information, visit \textit{http://astrodeep.eu }.}.
The detection has been performed on the HST F160W  image (\textit{H} band). We also add an additional set of sources detected in a stacked infrared (HST \textit{Y} + \textit{J} + \textit{JH} + \textit{H}) image, to recover additional faint sources. The final photometric catalogue consists of 10 bandpasses: HST ACS \textit{B}435, \textit{V}606, \textit{I}814; HST WFC3 \textit{Y}105, \textit{J}125, \textit{JH}140 and \textit{H}160; VLT Hawk-I \textit{Ks} 2.146 $\mu$m (ground based) and Spitzer IRAC 3.6 and 4.5 $\mu$m. 

To detect and measure the fluxes from the faint sources hidden behind the extended halos of the bright cluster galaxies, and within the IntraCluster Light (ICL), we developed an articulate procedure to process the two cluster images, with the goal of removing the foreground light letting the fainter and hidden sources behind it to appear. The parallel fields, on the other hands, did not require such an intervention and could be processed straightforwardly with the same approach used in the CANDELS campaign \citep[see][for a reference description of the adopted methods]{Galametz2013}.

The final catalogs and images are publicly available and can be downloaded from the \textsc{Astrodeep} website at \texttt{http://www.astrodeep.eu/frontier-fields/}; images and catalogues can also be browsed from a dedicated CDS interface at \texttt{http://astrodeep.u-strasbg.fr/ff/index.html}. A companion paper \citep[][ C16 from now on]{Castellano2016} presents the first scientific applications (photometric redshifts, magnification, physical properties). Throughout these works, AB magnitudes have been adopted.

The structure of this paper is as follows. Section \ref{dataset} describes the dataset we use in this study. Section\ref{detim} gives a detailed description of the procedure we applied to remove foreground light from the cluster detection \textit{H}160 image. In Section \ref{detcat} we describe how the detection catalogue is produced, and in Section \ref{HSTphot} the recipes used to obtain photometric measurements on the other HST bands are presented. Section \ref{KIRAC} presents the method we adopted to obtain the photometric measurements on the \textit{Ks} and IRAC bandpasses. Section \ref{IRdetection} describes an additional and complementary  detection process we perform on a stack of four infrared images. Section \ref {diagnostics} presents some diagnostics on the reliability of the results. Finally, Section \ref{conclusions} provides a summary of the work and a discussion on the results.


\section{The dataset} \label{dataset}

\begin{table*}[t!]
\begin{center}
\begin{tabular}{ | l || c | c || c | c | }
\hline
\textbf{Image} & PSF FWHM ('')& Limiting AB magnitude & PSF FWHM ('') & Limiting AB magnitude\\ \hline \hline
& \multicolumn{2}{|c||}{\textbf{A2744 Cluster}} & \multicolumn{2}{|c|}{\textbf{A2744 Parallel}}\\ \hline
ACS $B$435 & 0.11 & 28.58 &  0.12 & 28.97 \\ \hline
ACS $V$606 &  0.11 & 28.71 &  0.15 & 29.06 \\ \hline
ACS $I$814 &  0.13 & 29.03 & 0.13 & 29.17 \\ \hline
WFC3 $Y$105 &  0.18 & 29.15 & 0.19 & 29.30 \\ \hline
WFC3 $J$125 &  0.19 & 28.83 & 0.19 & 28.88 \\ \hline
WFC3 $JH$140 &  0.19 & 28.93 & 0.20 & 28.91 \\ \hline
WFC3 $H$160 &  0.20 & 29.11 & 0.20 & 29.06 \\ \hline
Hawk-I $Ks$ 2.146 & 0.38 & 26.13 & 0.38 &  26.12 \\ \hline
IRAC 3.6 &  1.66 &  24.83 & 1.66 & 24.83 \\ \hline
IRAC 4.5 &  1.72 & 24.87 & 1.72 & 24.87 \\ \hline
& \multicolumn{2}{|c||}{\textbf{M0416 Cluster}} & \multicolumn{2}{|c|}{\textbf{M0416 Parallel}} \\ \hline
ACS $B$435 &  0.12 & 28.86 &  0.13 & 28.81 \\ \hline
ACS $V$606 &  0.16 & 28.97 &  0.13 & 28.83 \\ \hline
ACS $I$814 &  0.16 & 29.31 &  0.18 & 29.19 \\ \hline
WFC3 $Y$105 & 0.18 & 29.22 & 0.18 & 29.28 \\ \hline
WFC3 $J$125 & 0.19 & 28.90 &  0.18 & 29.05 \\ \hline
WFC3 $JH$140 & 0.20 & 28.95 & 0.19 & 29.12 \\ \hline
WFC3 $H$160 &  0.20 & 29.01 & 0.20 & 29.14 \\ \hline
Hawk-I $Ks$ 2.146 & 0.38 & 26.18 & 0.38 & 26.26 \\ \hline
IRAC 3.6 & 1.66 & 25.04 & 1.66 & 25.11 \\ \hline
IRAC 4.5 & 1.72 & 25.05 & 1.72 & 24.89 \\ \hline \hline
\end{tabular}
\end{center}
\caption{PSF FWHM and depths of the dataset. Limiting magnitudes of HST images have been computed as the magnitudes within $2\times$ $FWHM_{H160}$ ($=0.2"$) circular apertures of $5\sigma$ sources, measured with \textsc{SExtractor} on PSF-matched images. Hawk-I and IRAC limiting magnitudes have been obtained computing $f_{5\sigma}=5 \times \sqrt{A_{aper}} \times f_{RMS}$ in each pixel, and taking the mode of the distributions ($A_{aper}$ is the area of a circular region with radius $r_{aper, K_s}=0.4", r_{aper, IRAC1}=1.66", r_{aper, IRAC2}=1.72"$).  } \label{d1}
\end{table*}

A2744 and M0416 image datasets are the first two publicly released of a total of six twin fields, observed by HST in parallel (i.e. the cluster pointing together with a ``blank'' parallel pointing) over two epochs, to a final depth of 140 orbits per field (FF program 13495, P.I. Lotz).
 The A2744 dataset also includes data acquired under programs 11689 (P.I. Dupke), 11386 (P.I. Rodney), and 13389 (P.I. Siana). The M0416 dataset combines the FF program data with imaging from the CLASH survey (P.I. Postman) and program 13386 (P.I. Rodney). The HST dataset consists of the following 3 optical and 4 near-infrared bands: $B$435, $V$606 and $I$814W (ACS);  $Y$105, $J$125, $JH$140 and $H$160 (WFC3).  We use the final reduced and calibrated v1.0 mosaics released by STScI, drizzled at  0.06" pixel-scale. A detailed description of the acquisition strategy and of the data reduction pipeline, as well as the final number of orbits in each band, can be found in the STScI data release documentation \footnote{https://archive.stsci.edu/pub/hlsp/frontier/}.  
 
 Additional information redward of WFC3 \textit{H}160 band is a fundamental  ingredient for selecting and investigating extragalactic sources across a  wide redshift range. We use the publicly released Hawk-I@VLT \textit{Ks} images of the A2744 and M0416 fields (P.I. Brammer, ESO Programme 092.A-0472\footnote{http://gbrammer.github.io/HAWKI-FF/}), resampled to the HST pixel scale using \textsc{Swarp} \citep{Bertin2002}. The final exposure time is 29.43 and 25.53 hours for A2744 and M0416, respectively. Finally, we include IRAC 3.6 and 4.5 $\mu$m data acquired under DD time and, in the case of M0416, Cycle-8 program iCLASH (80168, P.I. Capak). The final exposure time is $\sim$ 50 hours per field.

In Table \ref{d1} we list Point Spread Function (PSF) FWHM and depths (see Section \ref{HSTphot} and Section \ref{KIRAC}) of all the imaging data under investigation. To estimate the limiting magnitudes of Hawk-I and IRAC images, we use the RMS maps, corrected as explained in Section \ref{tphotprep}. The resulting values are in good agreement with those obtained by \citet{Laporte2015}.

\section{Preparing the clusters detection images}\label{detim}

\begin{figure*}[ht] 
\includegraphics[width=12cm]{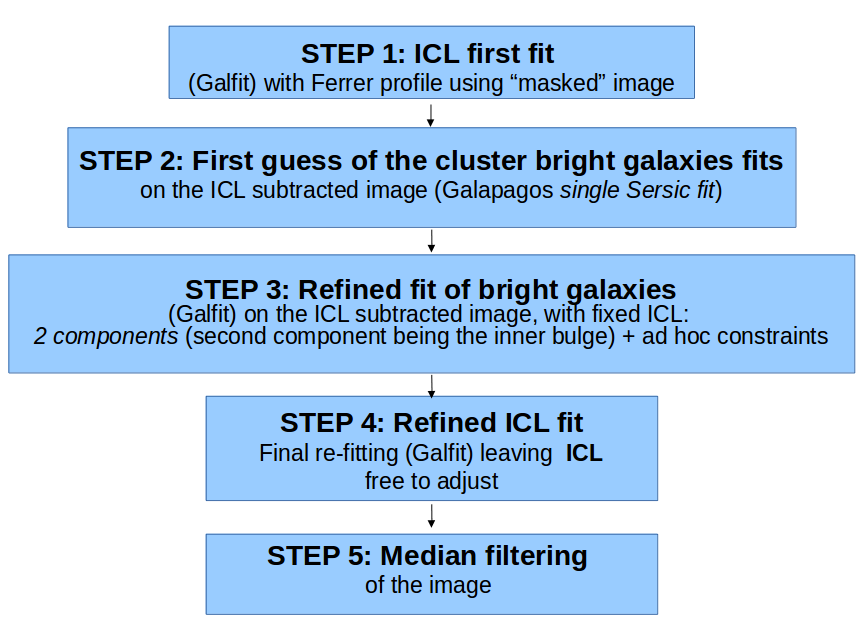}
\centering
\caption{Schematic description of the procedure applied to ``clean'' the cluster images removing the light from the foreground bright sources.}\label{method}
\end{figure*}

As anticipated, the goal of obtaining a deep and complete catalogue requires the development of a method to remove the light of foreground objects in the clusters fields. To this aim, the first step is performed on the image on which the detection of sources will be performed, i.e. the WFC3 \textit{F160} \textit{H}-band. As described in details later, we finally add a subset of sources detected on a different image (namely, a stack of infrared HST images), but the bulk of the detection is performed on the \textit{H} passband, as in the CANDELS and 3D-HST surveys, with which we choose to keep consistency. We therefore focus our attention on the \textit{H}-band detection process, leaving the description of the infrared stacking detection to Section \ref{IRdetection}.

Our basic strategy is to use the public code \textsc{Galfit} (Peng et al. 2010) to model as accurately as possible both the ICL and the brightest cluster members, and eventually subtract them from the images. We fine-tune our procedure by testing several possible variants. Experimenting multicomponent, multi-object fits with \textsc{Galfit} is a particularly tedious and time-consuming exercise, partly because of the large computing time required by \textsc{Galfit}, and mostly because of the numerical instabilities in its convergence. We have chosen to present in the paper the final solution that we adopted, that granted us the best result in the object subtraction. As shown below, it involves an iterative fitting of ICL and bright galaxies, in separate steps. Unfortunately, the conceptually most appealing solution (i.e. a simultaneous fit of both ICL and bright galaxies) did not produce comparably good results because of the degeneracy of allowed mathematical solutions to the problem, and has not been adopted here. 

The final product of the whole process is a residual image in which the ICL and the bright objects have been subtracted efficiently. It is important to stress that the goal of the procedure is not to obtain physically reasonable numbers for light profile slopes and magnitudes (although in most cases we do get sensible figures, which is undoubtedly reassuring, see Section \ref{diagnostics} and Fig. \ref{modelsSEDs}), but only to clean the image in the most efficient way, allowing for a more effective detection of faint sources. To obtain the most accurate models would require a tailored, ad-hoc fine tuning by hand of all the involved parameters and constraints \citep[as done e.g. by][]{Giallongo2014}. However, this would be beyond our aims of simply improving the catalogue assembly, and it would not be unsupervisedly repeatable on different fields. Therefore, we choose to stick to our simpler approach.

The basic steps we perform to remove the light of the cluster sources are as follows.

\begin{itemize}
\item \textbf{STEP 1}: using \textsc{Galfit}, we fit the diffuse ICL component after masking the brightest galaxies of the cluster, in order to obtain a first approximation model for the ICL, that we eventually subtract from the original image.
\item \textbf{STEP 2}: on the ICL-subtracted image, we use the public code \textsc{Galapagos} \citep{Barden2012} to obtain a first guess for the selected bright galaxies analytic profile.
\item \textbf{STEP 3}: again on the ICL-subtracted image, we use \textsc{Galfit} to refine the first \textsc{Galapagos} fit, including multiple components to better match the light profile in the central regions of the sources.
\item \textbf{STEP 4}: on the original image we let the ICL fit free to adjust, keeping frozen the fits for the galaxies\footnote{We point out here that attempts to let \emph{both} ICL and galaxies free to adjust in a final step gave unsatisfactory results, because of the degeneracy of allowed mathematical solutions to the problem. Fitting the ICL alone proved to provide a cleaner result.}.
\item \textbf{STEP 5}: we eventually apply a median filter on the residual images that removes the remaining intermediate scale background residuals.
\end{itemize}

The global method is sketched in the flowchart in Fig. \ref{method}.
In the following we describe in more details each step of the procedure.

\subsection{ICL first guess fit}

To obtain a reasonable first guess model for the ICL we first crudely mask out the contribution of the brightest sources. To do so, we start estimating the $1\sigma_{\rm sky}$ level of the \textit{H}-band image by measuring $\sigma$ in relatively small and blank boxes distributed homogeneously in regions sufficiently far from the cluster core; then we produce a mask image from the \textit{H}-band image where all pixels above $8\sigma_{\rm sky}$ will be excluded from the \textsc{Galfit} fit. This cut masks out most of the light from the brightest sources in the core of the cluster, leaving enough pixels to fit the ICL with \textsc{Galfit}.

In order to do so, we also need information on the image noise, and its PSF. We let \textsc{Galfit} create a noise image internally, using the \texttt{gain}, \texttt{exptime} and \texttt{rdnoise} header parameters of the science image. On the other hand, we create a PSF model by stacking the cutouts of isolated, bright, but not saturated stars in the field, selected with an ad-hoc \textsc{SExtractor} routine (we use version 2.8.6 in this work).

Then, following \citet{Giallongo2014}, we use \textsc{Galfit} (version 3.0.2) to fit a modified Ferrer profile \citep{Binney1987},
\begin{equation}
 \Sigma = \Sigma_0 \ (1 - (r/r_{\rm out})^{2-\beta})^{\alpha} \label{Ferrer}
\end{equation}
to the diffuse ICL distribution to all pixels with non-zero values. The Ferrer profile has nearly flat core and an outer truncation set by $r_{\rm out}$, being similar in shape to a S\'ersic profile \citep{Sersic1968} of index $n<0.5$ (see next Section). First guesses for $\Sigma_0$,  $r_{\rm out}$, $\beta$ and $\alpha$ were estimated from the 1D isophotal profile. 

The initial best-fit model for the ICL is intended to describe the overall shape of the ICL profile. For this reason, in addition to the Ferrer profile, we leave the position angle, the axis ratio and the boxiness/diskyness parameter of the models free to vary. In addition, for A2744 we add one bending mode ($B2$) to follow more closely the global shape of the ICL isophotes. For M0416, instead, we find our best fit using two Ferrer components, without bending modes.

\subsection{Bright galaxies first guess fit}

To fit the bright foreground galaxies belonging to the clusters, we follow an iterative procedure. \textsc{Galfit} is a very powerful tool, but can be prone to degeneracy instabilities when dealing with multi-component fitting. To avoid the risk of misinterpreting local minima as best fits, we proceed by steps of increasing complexity, first fitting the objects with a single S\'ersic profile, 
\begin{equation}
\Sigma(R)=\Sigma_e\exp\left\lbrace-b_n\left[\left(\frac{r}{r_e}\right)^{1/n_{Sersic}}-1\right]\right\rbrace, \label{Sersic}
\end{equation}
and then refining the fit adding secondary components, as described in more details in the following Sections.

The first step is performed by means of a \textsc{Galapagos} run on the ICL-subtracted image. \textsc{Galapagos} is a software package which directly links  \textsc{SExtractor}  and \textsc{Galfit}, allowing for a smooth dataflow from the detection of sources to the estimation of their analytical fit. It is extremely useful for automatic and simultaneous fitting of many objects, removing also fainter contaminants, that are of course very numerous in our case.

To identify the bright objects we want to fit, we take advantage of the first stage of the \textsc{Galapagos} pipeline, which consists of a \textsc{SExtractor} run to detect sources and measure the background. We select galaxies with \texttt{MAG\_AUTO} $<19$, retrieving 15 objects in the case of A2744 and 20 in the case of M0416.

We then run the subsequent \textsc{Galapagos} stages, automatically fitting the selected sources with single-S\'ersic analytical profiles (the fit is performed including the neighbouring galaxies and the local background). For each object, we obtain an analytical model giving the best estimation of the total magnitude, the effective radius $r_e$, the S\'ersic index $n_{Sersic}$, the axis ratio and the sky-projected position angle.

\subsection{Bright galaxies refinement fit}

We then proceed to the refinement stage, again on the ICL subtracted image, directly using \textsc{Galfit} with the output parameters from the \textsc{Galapagos} fits as starting guesses. We introduce a second component, to fit the central regions of the objects more accurately. In practice, we include a ``bulgy'' component, with $n_{Sersic}>4$ and axis ratio $q>0.5$, alongside a generally more ``disky'' component, which we force to have $n_{Sersic}<4.5$\footnote{Of course, a Sersic index $n_{Sersic} \sim 4$ can hardly describe a disk; here we use this terminology for the sake of conciseness.}.
In this stage, all fitting parameters are let free to vary, with constraints on positions ($\pm$1 pixels for the disky component, $\pm$3 for the bulgy component), magnitudes ($\pm$1 for the disky component, -1/+5 for the bulgy component), effective radii (1 to 60 pixels for the disky component, 1 to 30 for the bulgy component), S\'ersic index (0.5 to 4.5 for the disky component, 4 to 8 for the bulgy component), and axis ratio (0.5 to 1 for the bulgy component).

When necessary, we further refine the fit with additional loops to add more components (for example, in the case of a very bright and centrally concentrated source in M0416, for which a PSF-shaped source has been added to the two main components).

\subsection{ICL refinement fit}

We finally run again \textsc{Galfit} on the original, pre-ICL subtraction image, fixing all galaxies parameters and adding the ICL model(s). The ICL centroid position, the central surface brightness and the truncation radius are let free to adjust (with constraints $\pm$1 pixel, 25 to 27, and 1000 to 8000 pixels, respectively). Reassuringly, the final fit does not dramatically diverge from the first guess: the central surface brightness $\Sigma_0$ changes from 25.8957 to 25.3857 for A2744 and from 25.50 to 25.5338 (first component) and from 25.5000 to 25.8423 (second component) for M0416. Radii and morphological shapes undergo similar minor changes as well.

\subsection{Median filtering} \label{medianfilter}

\begin{figure*}[ht] 
\includegraphics[width=16cm]{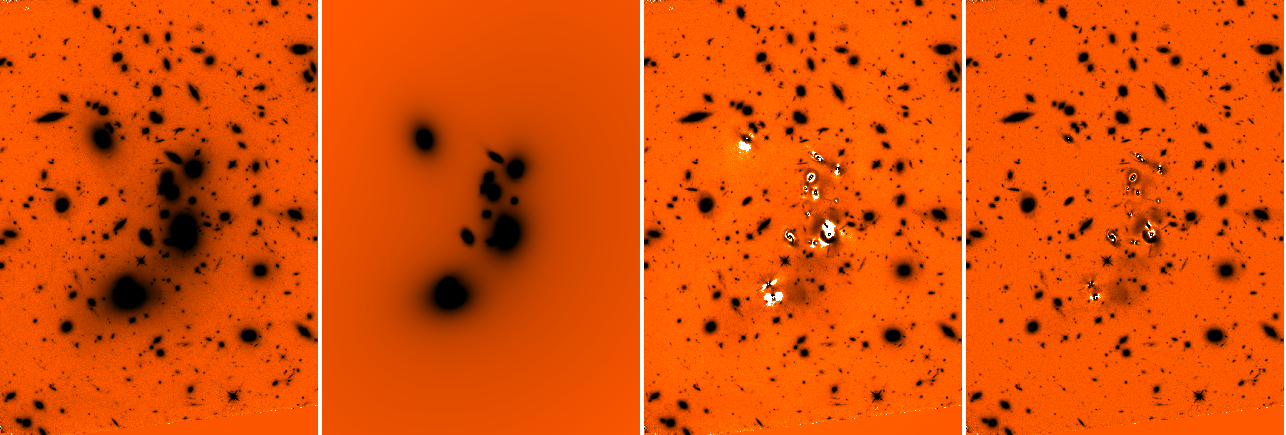}\vspace{2mm}
\includegraphics[width=16cm]{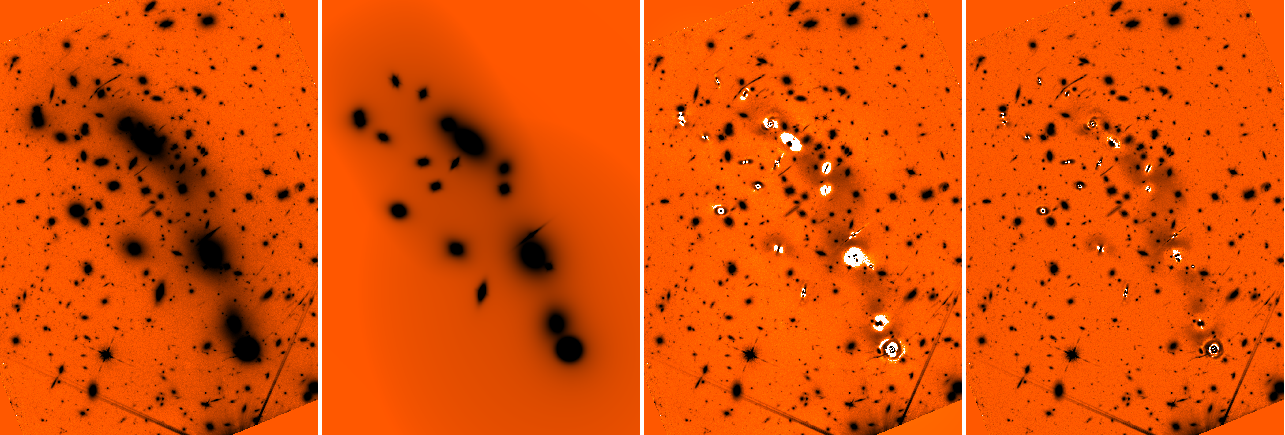}
\centering
\caption{Effects of the procedure on the A2744 (top) and M0416 (bottom) \textit{H}160 image. Left to right: original image, models of galaxies and ICL after STEP 4, processed image after subtraction of the models, final processed image after median filtering (STEP 5). All images have logarithmic greyscale with the same cuts.}\label{proc_all}
\end{figure*}

The final step of our procedure aims at further improving the detection and the photometry in regions close to bright galaxy centers, by mitigating the effects of small scale negative \textsc{Galfit} residuals, that could hamper local background estimates. We use the IRAF task \texttt{median} to create median filtered version of the residual images over a 1"$\times$1" box. To avoid affecting low signal-to-noise pixels belonging to the detected sources, we exclude from the computation all pixels at $>$1$\sigma_{SEx}$ above zero counts (where $\sigma_{SEx}$ is computed by \textsc{SExtractor}), and their nearest neighbours. The resulting median-filtered image is then subtracted from the original one, obtaining a clear improvement in the innermost regions of the cluster. A comparison of the photometry extracted from the original residual image and the median-subtracted one shows no significant difference for any source far from the cluster center, demonstrating that this last step is effective at improving the processed images while leaving flux measurements unaltered; this is further confirmed from simulations tests (see Section \ref{completeness}) showing that measured fluxes of test fake sources are less scattered with respect to their true input fluxes when performing measurement on the median-subtracted image. 
Fig. \ref{proc_all} shows the effects of the procedure on the two cluster fields.

\subsection{Corrected RMS map}

As a final refining step, we use the \textsc{Galfit} model image to estimate, on the basis of the image exposure time, the photon noise in each pixel contributed by the bright \textsc{Galfit}-subtracted sources.  The resulting ``photon noise image'' is then added to the original RMS map, to take into account the effect of the aforementioned subtracted sources on the detection and the flux measurement in the innermost cluster regions. 

\section{Obtaining the detection catalogues} \label{detcat}

\subsection{Detection strategy} \label{detstr}

To obtain the final detection catalogues on the \textit{H} processed images we use \textsc{SExtractor} with a \texttt{HOT+COLD} approach \citep{Galametz2013}. This procedure adopts two different sets of the \textsc{SExtractor} parameters to detect objects at different spatial scale. The COLD mode is used to detect bright extended sources, with relatively low efficiency in deblending. The HOT mode is more efficient for faint galaxies and is used to detect small sources outside the region of the brightest sources that are already detected with the COLD mode. Typically, the COLD mode is used to detect sources that are ~1.5 mag brighter than the detection limit. We remark that in both case we refer to objects that are however fainter than the brighter cluster members that we have already fitted and removed.
Considering that in the clusters fields, even after the described procedure helped cleaning out the image, many faint sources lie within or nearby extended halos of bright galaxies, we find more convenient to adopt a more aggressive set of parameters, so that our \texttt{COLD} detection actually corresponds to the CANDELS \texttt{HOT} detection. Very faint sources are then detected using an even more extreme choice of the parameters, so that our \texttt{HOT} detection turns out to be ``hotter'' than the CANDELS one, with an aggressive choice for the background subtraction. We choose to use the same recipe on the parallel fields as well. We list the relevant parameters adopted in this procedure in the Appendix.

\begin{figure*}[ht] 
\includegraphics[width=11cm]{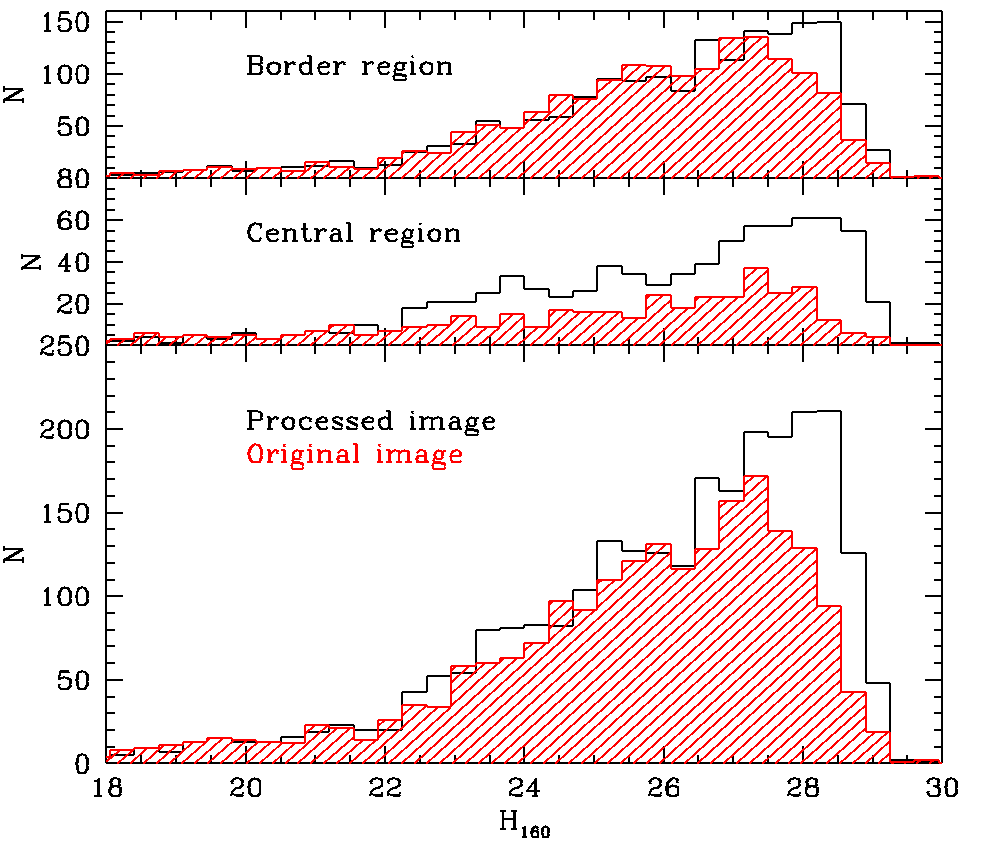}
\centering
\caption{Detected objects on A2744 cluster field with and without the procedure. Top: border region; middle: central region; bottom: all field. The central region is defined as a square area centered on the cluster with size 1200 pixels.}\label{countsprocedure}
\end{figure*}

\begin{figure*}[ht] 
\includegraphics[width=16cm]{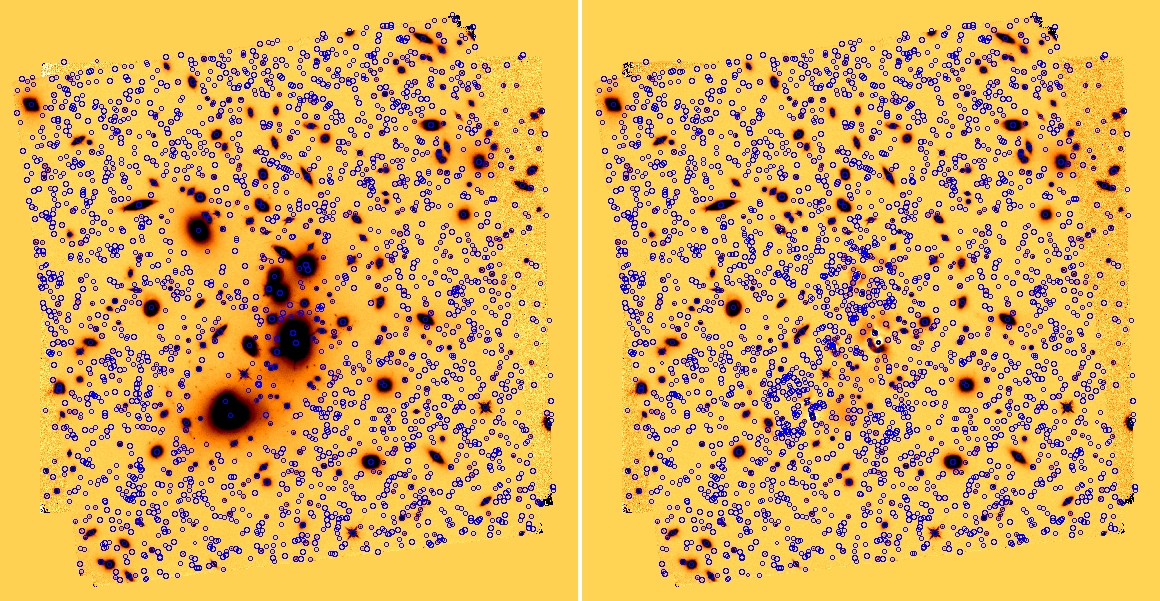}
\centering
\caption{Detected objects on A2744 cluster field with (right) and without (left) the procedure.}\label{confronto}
\end{figure*}

Table \ref{tabdetect} lists the total number of detected sources on the $H$160 images in the four considered fields. It is interesting to check how much our procedure to remove ICL and bright cluster galaxies improves the detection of faint sources. At this purpose we compare the number of detected sources in the A2744 cluster field with respect to a \texttt{HOT+COLD} \textsc{SExtractor} run on the original \textit{H}-band image (using identical parameter settings for the detection). The results are shown in Figs. \ref{countsprocedure} and \ref{confronto}. In the central regions, we are able to increase the number of detected sources by almost a factor of two, as shown in the mid panel of Fig. \ref{countsprocedure}. Looking at the right panel of Fig. \ref{confronto}, it can be seen that many detected sources lie close to the removed cluster members. Since it is possible that at least some of them are spurious detections, we flagged them in the final \textsc{SExtractor} catalogue, to keep track of potential flaws in the following stages of the process. We choose to flag any detected source having its centroid lying in a region where the normalized flux per pixel of a bright source model is above a given threshold $f_{flag}$. We empirically find that taking $f_{flag}=0.1$ yields reasonable results. See Appendix \ref{catformat} for more details on this flag.

\begin{table}[t!]
\begin{center}
\begin{tabular}{ | l | c || l | c | }
\hline
\textbf{Image} & $N_{detect}$ & \textbf{Image} & $N_{detect}$ \\ \hline
A2744$_{cl}$ & 2596 (+15) & M0416$_{cl}$ & 2556 (+20) \\ \hline
A2744$_{par}$ & 2325 & M0416$_{par}$ & 2581 \\ \hline
\end{tabular}
\end{center}
\caption{Total number of detected sources on the $H$160 images in the four considered fields. The numbers among brackets are the models of cluster bright objetcs (15 for A2744$_{cl}$, 20 for M0416$_{cl}$).} \label{tabdetect}
\end{table}
 
\subsection{Completeness} \label{completeness}

\begin{figure*}
\centering
\includegraphics[width=8cm]{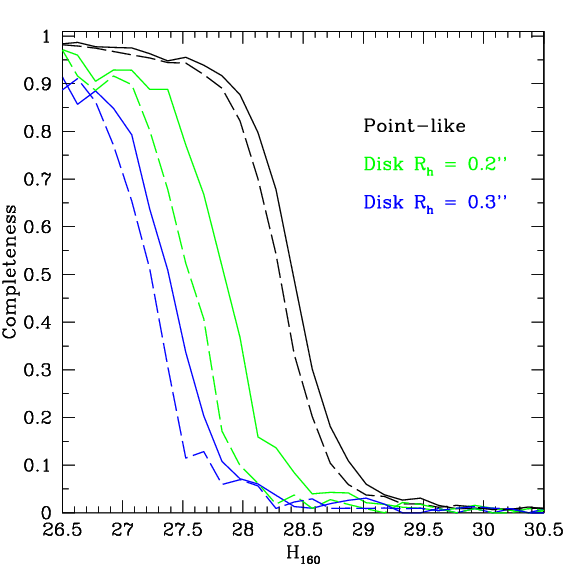}
\caption{Detection efficiency (H-detected catalogue) for point-like and disk-like sources in A2744 (continuous lines) and M0416 (dashed lines).} \label{completeness}
\end{figure*}
    
We assess detection completeness as a function of the \textit{H}-band magnitude by running simulations with synthetic sources. Using in-house developed scripts, we first generate populations of point-like (i.e. PSF shaped) and exponential profile sources, with total \textit{H}-band magnitude in the range 26.5-30. Disc-like sources are assigned an input half-light radius R$_{h}$ randomly drawn from a uniform distribution between 0.0 and 1.0 arcsec. These fake galaxies are placed at random positions in our detection image, avoiding positions where real sources are observed on the basis of the original \textsc{SExtractor} segmentation map.
To avoid an excessive and unphysical crowding of simulated objects, we include 200 objects of the same flux and morphology each time. We then perform the detection on the simulated image, using the same \textsc{SExtractor} parameters adopted in the real case. In Fig. \ref{completeness} we show the completeness (ratio of the number of recovered to input sources) as a function of the total input magnitude of the simulated objects. 
We find that the 90\% detection completeness for point sources is at \textit{H}$\sim$27.7-27.8, and decreases to \textit{H}$\sim$27.1-27.3 and \textit{H} $\sim$26.6-26.7 for disk-like galaxies of R$_{h}$=0.2 arcsec and R$_{h}$=0.3 arcsec respectively, the lower values referring to M0416 as expected from its slightly shallower \textit{H}-band depth.

\section{Photometry on the HST images} \label{HSTphot}

Having obtained the detection catalogue on the HST \textit{H}-band images, we proceed to perform the photometric measurements on the other HST bands. To this aim, we process each cluster image to remove foreground sources and ICL, as for the \textit{H} band cluster images (again, this is not necessary for the parallel field images). However, having already obtained a robust two component Sersic fit in the \textit{H} band, now we can adopt a simpler approach: for each HST image, we use the output from the final \textsc{Galfit} run on the closest redder band as an input starting guess (e.g., \textit{H}160 output as input for \textit{JH}140, \textit{JH}140 output as input for \textit{J}125, etc.), letting the estimates for positions, magnitudes, radii and S\'ersic indexes free to vary (again applying reasonable constraints on the allowed range of variations: $\pm$3, -1/+5, 1 to 60 and 0.5 to 4.5 for disky components,  $\pm$3, -1/+5, 1 to 30, 4.0 to 8.0, and also imposing an axis ratio $q>0.5$ for the bulgy component). For the ICL, we impose $\Sigma_0$ -1/+5, radius 2000 to 5000.

After this procedure, we convolve all the images down to the \textit{H}160 resolution (FWHM=0.2"), with a convolution kernel obtained taking the ratio of the PSFs of the two images in the Fourier space. We also applied to all the images the median filtering process described in Section \ref{medianfilter}, since the simulations showed that the photometric measurements are more robust and the scatter in the measured fluxes is reduced.

We then run \textsc{SExtractor} in dual mode, to measure aperture and isophotal photometry. 

\section{K and IRAC photometry} \label{KIRAC}

As described in Section \ref{dataset}, several independent programs are obtaining data on the FF regions. In particular, both new deep \textit{K} and Spitzer (3.6 and 4.5 $\mu$m) are available for analysis (see Section \ref{dataset}).

To extract photometric measurements on such lower resolution NIR images, we perform a fit on the images with a PSF-matching technique, by means of the code \textsc{t-phot} \citep{Merlin2015}. \textsc{t-phot} uses the spatial and morphological information from a high-resolution image as priors to construct low-resolution normalized models, obtained via convolution with a PSF-matching kernel, and simultaneously fits all the objects in the field re-scaling the fluxes of such models. The code allows to use “real” cut-outs of sources together with analytical models as high-resolution priors; we took advantage of this option to simultaneously fit the faint sources with the models of the bright cluster members (the latter having the two components stacked to obtain single component models, in order to avoid possible degeneracy issues in the fitting procedure).\footnote{We first tried to process the measurement images with the same procedure adopted for the HST images, subtracting the foreground sources to fit only the \textit{H}-detected objects, but the results were not satisfactory.}

In order to minimize the effects of too small a segmentation, the \textsc{SExtractor} output map has been \textit{dilated} before being fed to \textsc{t-phot}, enlarging the size of the segmented area of each source \citep[the procedure is the same that has been adopted for the CANDELS \textsc{tfit} photometry and is described in][]{Galametz2013}.

\subsection{Preparation of the measurement images}\label{tphotprep}

We estimate the PSF on the Hawk-I \textit{K} images stacking well isolated stellar sources, with the same procedure adopted for the HST images. For IRAC, we find that a better approximation can be obtained using a synthetic PSF, constructed using a STScI script (kindly provided by H. Ferguson and S. Lee, private comm.) which stacks multiple renditions of the synthetic ideal PSF, each one oriented consistently with the position angle of a single pointing and weighted by the pointing exposure time.

We also correct the RMS maps, so that sources at the detection limit have measured fluxes distributed consintently with the statistical expectation. To do so, we randomly select 200 positions in the parallel fields, far enough from detected sources (this was achieved building a $2 \sigma$ level mask and considering only the unmasked regions). Then, we measure the flux of fake point sources injected at the selected positions, and compute the RMS map multiplicative factor required to make the distribution of the measured $S/N$ have standard deviation consistent with 1.

We also further correct the background, measuring the shift of the mean of the distribution of the same fake sources on copies of the images having small constant artificial background offsets, and computing the offset $\Delta_{bkgd}$ required to make the measured shift consistent with zero:
\begin{equation}
\Delta_{bkgd} = \frac{f_- + f_+}{f_- - f_+}\times \Delta_{bkgd,in} ,
\end{equation}
\noindent where $f_-$ and $f_+$ are the mean measured fluxes of the fake sources in two images having $\pm \Delta_{bkgd,in}$ small constant background offset. We finally assume that the background and RMS correcting factors obtained for the parallel fields are also valid for the clusters fields, since it would be hard to find a sufficient number of void regions in the crowded and relatively small cluster images.

\begin{figure*}[ht]
\centering
\includegraphics[width=14cm]{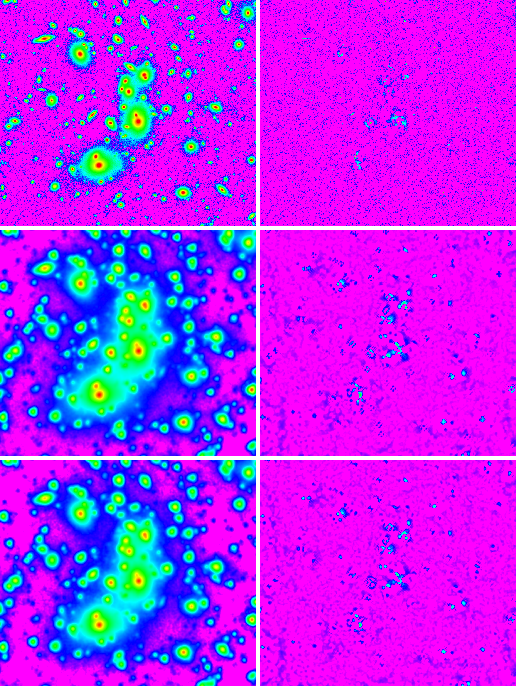}
\caption{A2744 residuals after \textsc{t-phot} processing. Left to right, top to bottom: original and residual images of K, IRAC-CH1 and IRAC-CH2, in logarithmic scale.} \label{A2744TPHOTres}
\end{figure*}

\begin{figure*}[ht]
\centering
\includegraphics[width=10cm]{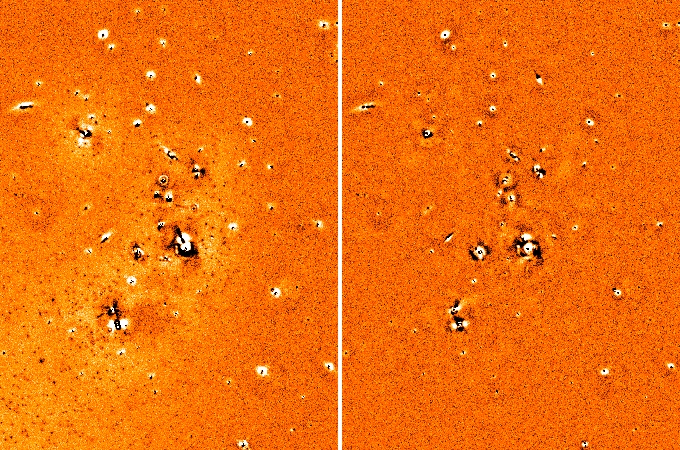}
\caption{A2744 \textit{Ks} residuals. Left: straightforward \textsc{t-phot} run; right: with local background subtraction procedure.} \label{A2744tphotbksub}
\end{figure*}

\subsection{T-PHOT runs}

After preparing the measurement images, we proceed with the \textsc{t-phot} runs. For the cluster images we found that simply including the ICL models in the list of sources and fitting it together with all the other priors yields unsatisfactory results. We have therefore developed a different procedure, where we estimate the local background independently for each source, and then combine all these measurements to build a large scale background image. To this purpose, we have included in the \textsc{t-phot} runs a more complex description of all objects, adding a free parameter to describe the local background. This has been possible without modifying the code simply adding and additional analytic component:
\begin{itemize}
\item for each source, a ``twin'' template with same extension and top-hat flat normalized profile has been coupled to the ``true'' model template (excluding the bright sources models);
\item these ``background'' templates are fitted together with the convolved models of the real cutouts, during the \textsc{t-phot} run;
\item a model image is produced using only the ``background'' templates, multiplied by their fitting factors;
\item this raw background map is then smoothed with a gaussian kernel, and subtracted from the original LRI.
\end{itemize}

The resulting images have the major fraction of the local background well removed. We finally fit these background subtracted images again with \textsc{t-phot}. Fig. \ref{A2744TPHOTres} shows the original images and the relative residual images, obtained subtracting the scaled models generated by \textsc{t-phot}. Fig. \ref{A2744tphotbksub} show the effects of the local background subtraction process in the case of the A2744 \textit{Ks} image, comparing the residuals from a straightforward \textsc{t-phot} run with the residuals obtained with our process.

We follow the same strategy to process the parallel fields (of course without the need to include any analytical model in the priors list).

\begin{figure}[t!]
\centering
\includegraphics[width=9cm]{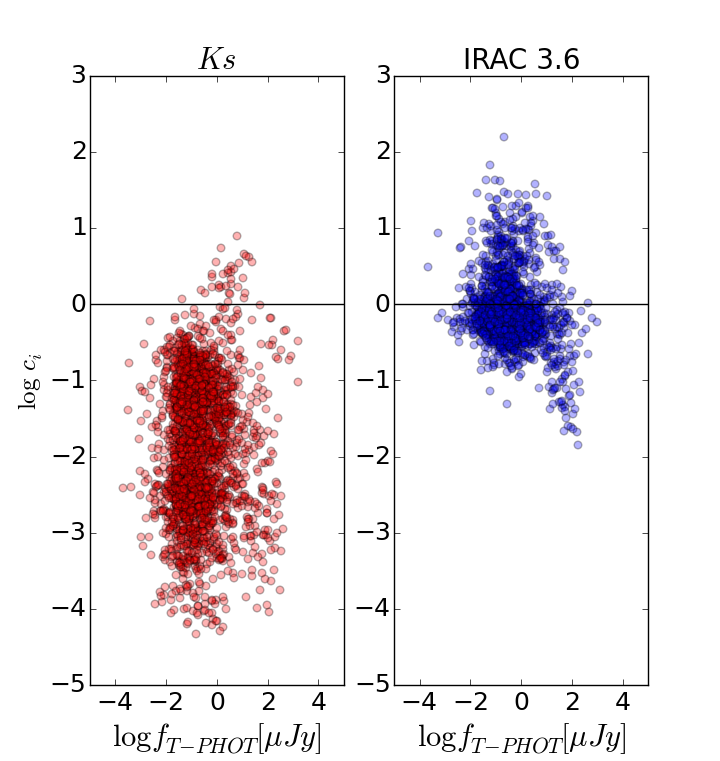}
\caption{Covariance index $c_i$ (defined for each source as the ratio between the maximum covariance and the variance) as a function of the measured flux using \textsc{t-phot} (shown in $\log \mu Jy$), in the A2744 cluster field. Left: $Ks$; right: IRAC $3.6 \mu m$. The horizontal line, at $\log c_i=0.0$, indicates a reference value of reliability for the measured flux. See text for more details.} \label{cov}
\end{figure}

In all cases, the uncertainties on the measured fluxes are given by \textsc{t-phot} as the square root of the diagonal terms in the covariance matrix of the system \citep[see][]{Merlin2015}. Additional information on the reliability of the fit are given by a set of flags in the \textsc{t-phot} output. In particular, the \textit{covariance index} $c_i$ (defined for each source as the ratio between its maximum covariance term and its variance, in the covariance matrix) is an important diagnostic that can point out unreliable measurements, helping to identify sources affected by strong contamination from neighbors, potentially causing failures of the measurement method. Fig. \ref{cov} shows the values of the covariance index as a function of the measured fluxes, for the cases of $Ks$ and IRAC $3.6 \mu m$ A2744 cluster field. Clearly, many IRAC sources show a high degree of contamination, which can (but not necessarily does) cause problems in the measurement. Sources having $c_i>1$ should be considered with caution. This diagnostics can be very useful for the correct determination of photometric redshifts, and we use it in our analysis see C16.

\section{Additional faint IR-detected sample} \label{IRdetection}

Performing the detection in a single band provides clear advantages in terms of a selection function that is more robust and easier to estimate. Our procedure to ``clean'' the image from bright cluster sources and from the ICL has allowed us to turn the FF cluster pointings into ``blank'' fields, such that our \textit{H}-band detected catalogue can be considered as complementary to public \textit{H}-detected catalogues in wider albeit shallower areas \citep[e.g.][]{Guo2013, Galametz2013}. However, the resulting catalog is not as deep as the one that could be obtained out of a stack of the IR images. For instance, a combined \textit{Y+J+H} detection is in principle more effective in selecting blue galaxies at redshifts $\sim 6-8$.
For this reason we complement our catalogues with lists of objects detected in a weighted mean of the \textit{Y}105+\textit{J}125+\textit{JH}140+\textit{H}160 bands while undetected in the \textit{H}-band. This IR-stack is built from the processed \textsc{Galfit} residual images, and used as detection band in the same way as the \textit{H}160 one. We then individuate all sources whose segmentation does not overlap with any pixel belonging to \textit{H}-detected sources according to the relevant \textit{H}160 segmentation map. In this way, we isolate 976, 1086, 832, 1152 objects in A2744 cluster, A2744 parallel, M0416 cluster and M0416 parallel respectively. These are mostly sources with S/N(H160)$<$5 and H160$\sim$27-30 (Fig.\ref{rawcounts}).
 
To assess consistency between the photometry in the \textit{H}-detected and the IR-detected catalogues we compare the fluxes of bright sources in common among the two (matched within 0.2" radius). We verify that the photometry is consistent with no appreciable offsets both in the HST and in the \textsc{t-phot} extracted bands. This test shows that no systematic is introduced in HST and in \textsc{t-phot} fluxes by the use of a different detection image, or by a different number of (faint) priors.


\section{Diagnostics} \label{diagnostics}

As a first check on the reliability of our results, we present here some diagnostic checks. 

  \begin{figure*}
   \centering
   \includegraphics[width=8cm]{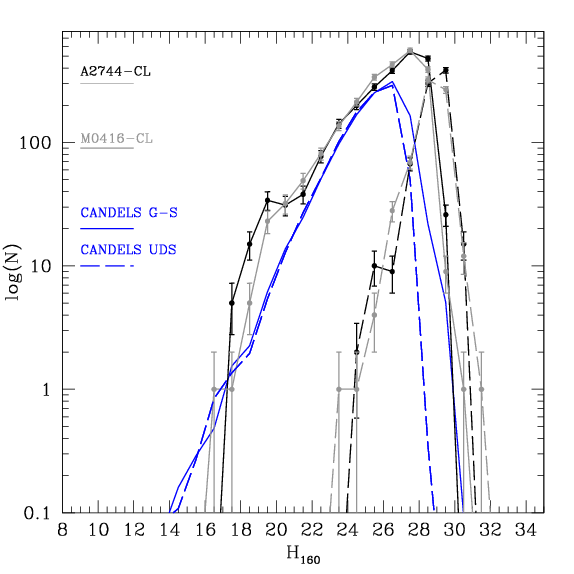}
   \includegraphics[width=8cm]{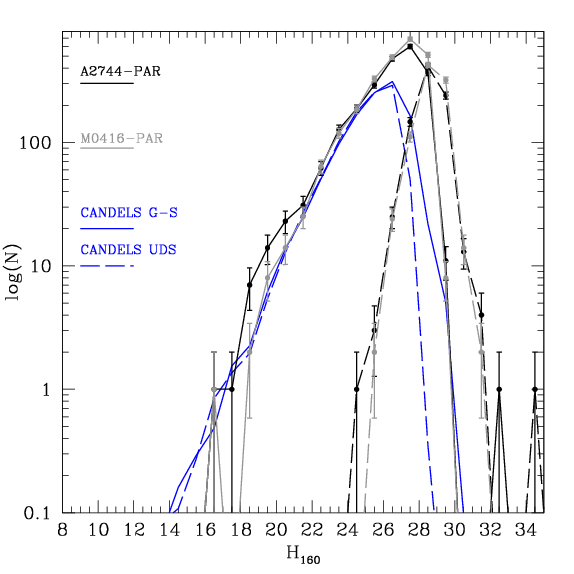}
   \centering
   \caption{Raw number counts of detected objects on the detection images. Top panel: A2744; bottom panel: M0416. Black lines refer to the cluster fields, blue lines to the parallel fields; solid lines refer to the $H$ detected catalogue, dashed lines to the additional IR-stack detected catalogue. For reference, the red lines refer to CANDELS fields number counts re-scaled to the FF area (solid: GOODS-South, dashed: UDS). See text for details.}
   \label{rawcounts}
\end{figure*}    

  \begin{figure*}
   \centering
   \includegraphics[width=8cm]{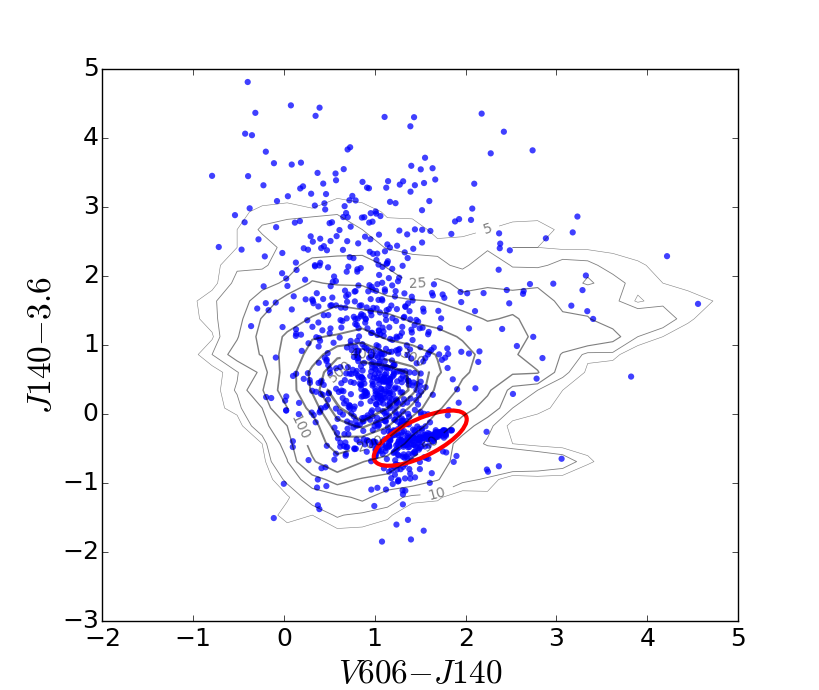}
   \includegraphics[width=8cm]{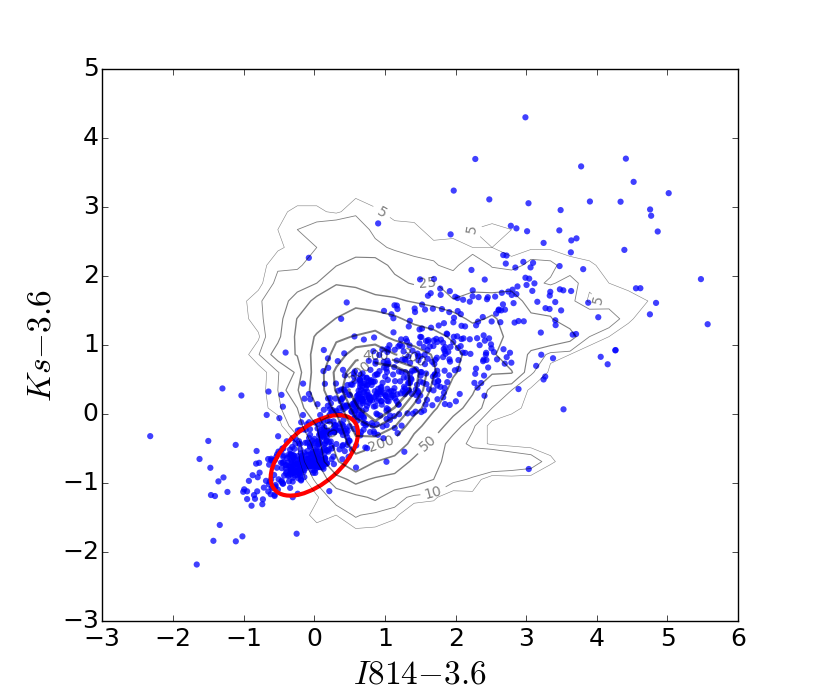}
    \includegraphics[width=8cm]{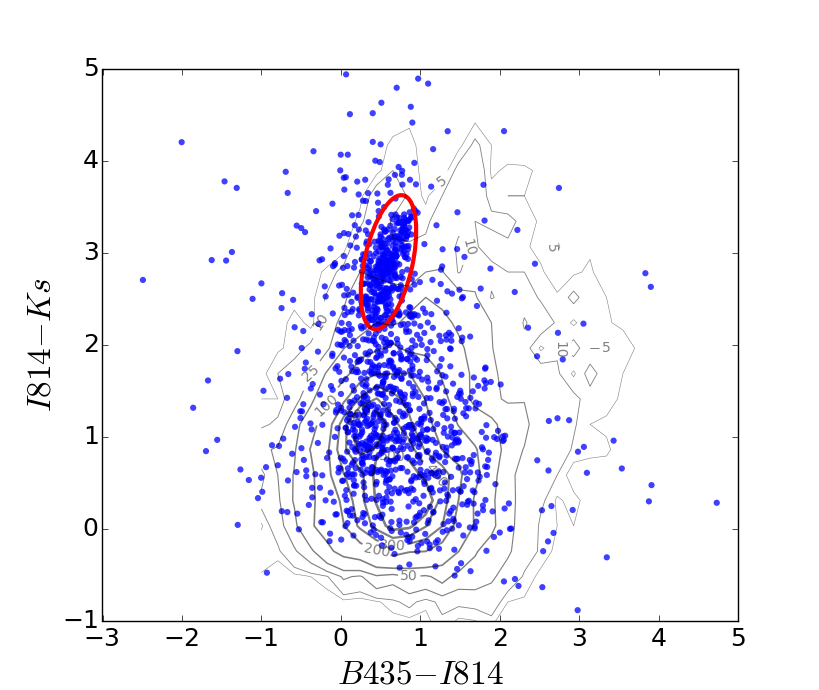}
   \centering
   \caption{Color-color diagrams. Grey lines are number density contours for GOODS-South CANDELS objects, using \citet{Guo2013} photometry. Blue dots: A2744 cluster field, this work (only objects detected in all relevant bands are included). The large majority of the objects inside the red ellipses, which deviate from the typical distributions, are likely to be cluster members, as confirmed by the photo-$z$ analysis in C16.}
   \label{colors}
\end{figure*}

  \begin{figure*}
   \centering
   \includegraphics[width=5cm,height=5cm]{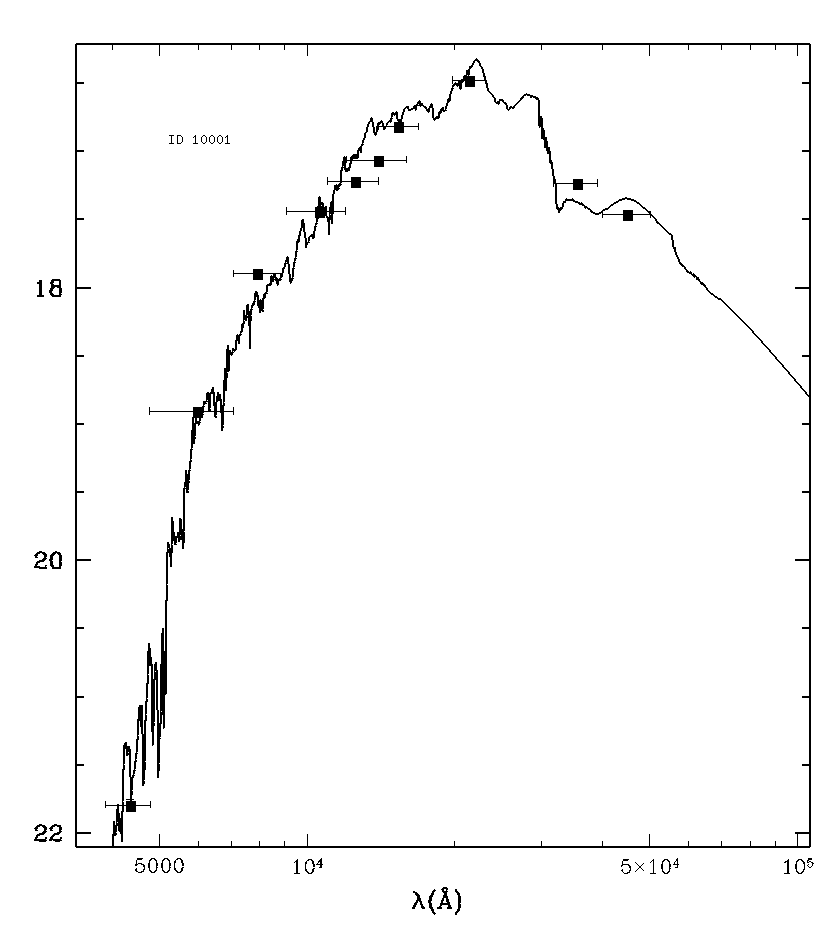}
   \includegraphics[width=5cm,height=5cm]{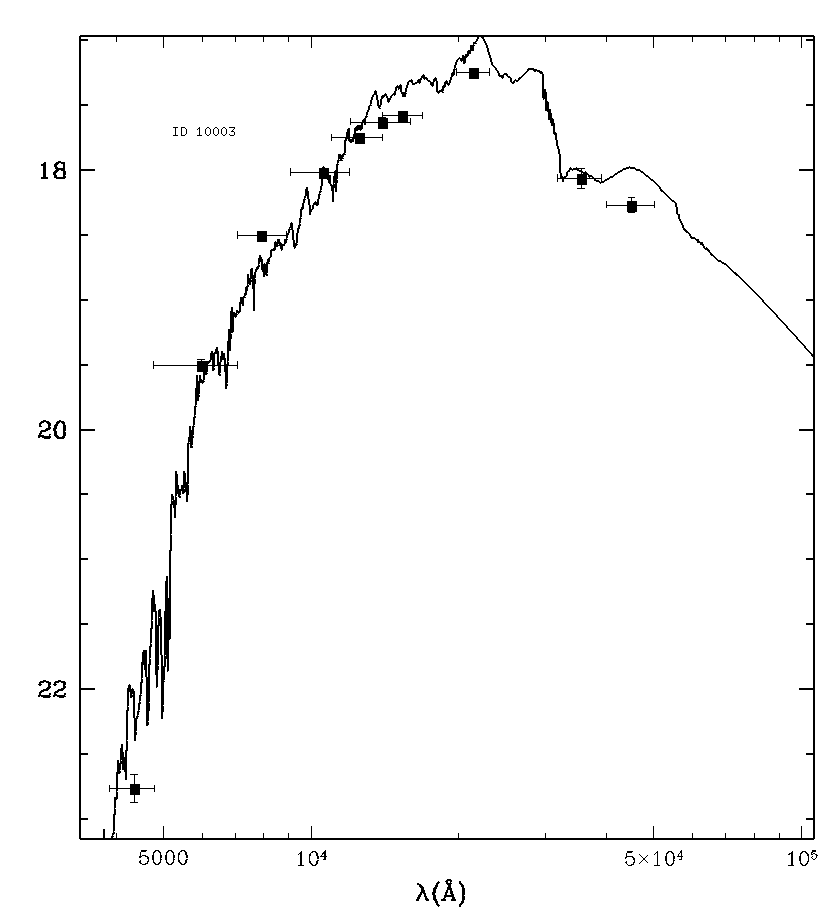}
    \includegraphics[width=5cm,height=5cm]{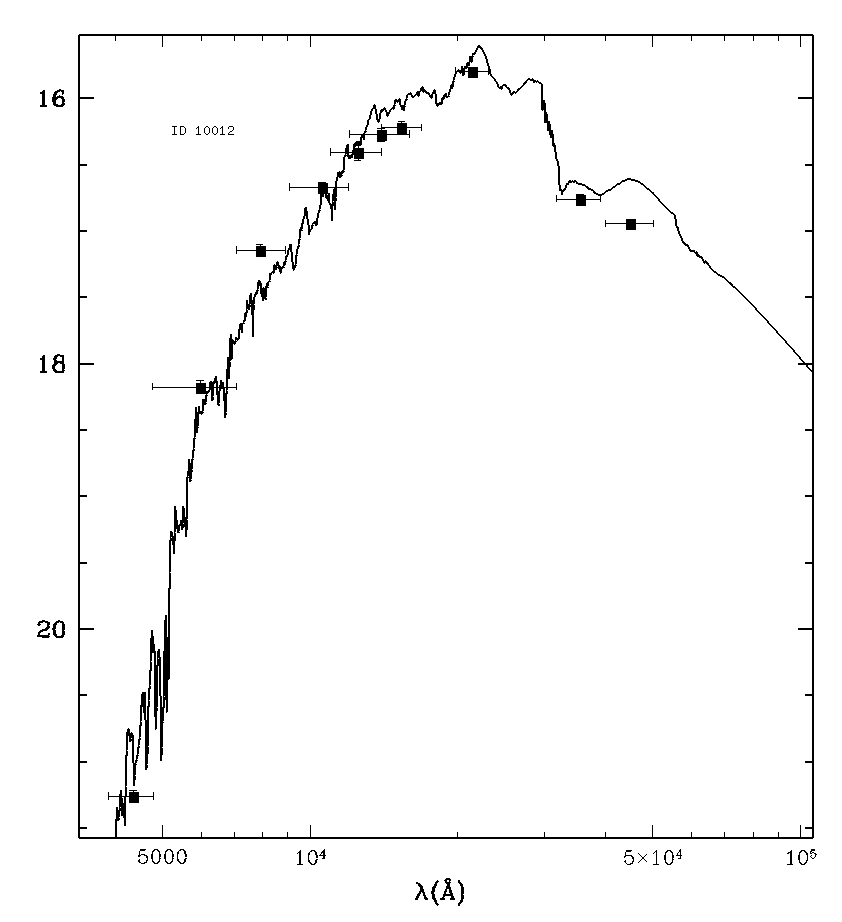}
   \centering
   \caption{Examples of SED fitting with photometry from this work, for three models of bright cluster sources belonging to A2744 ($z=0.308$).}
   \label{modelsSEDs}
\end{figure*}

  \begin{figure*}
   \centering
   \includegraphics[width=7cm]{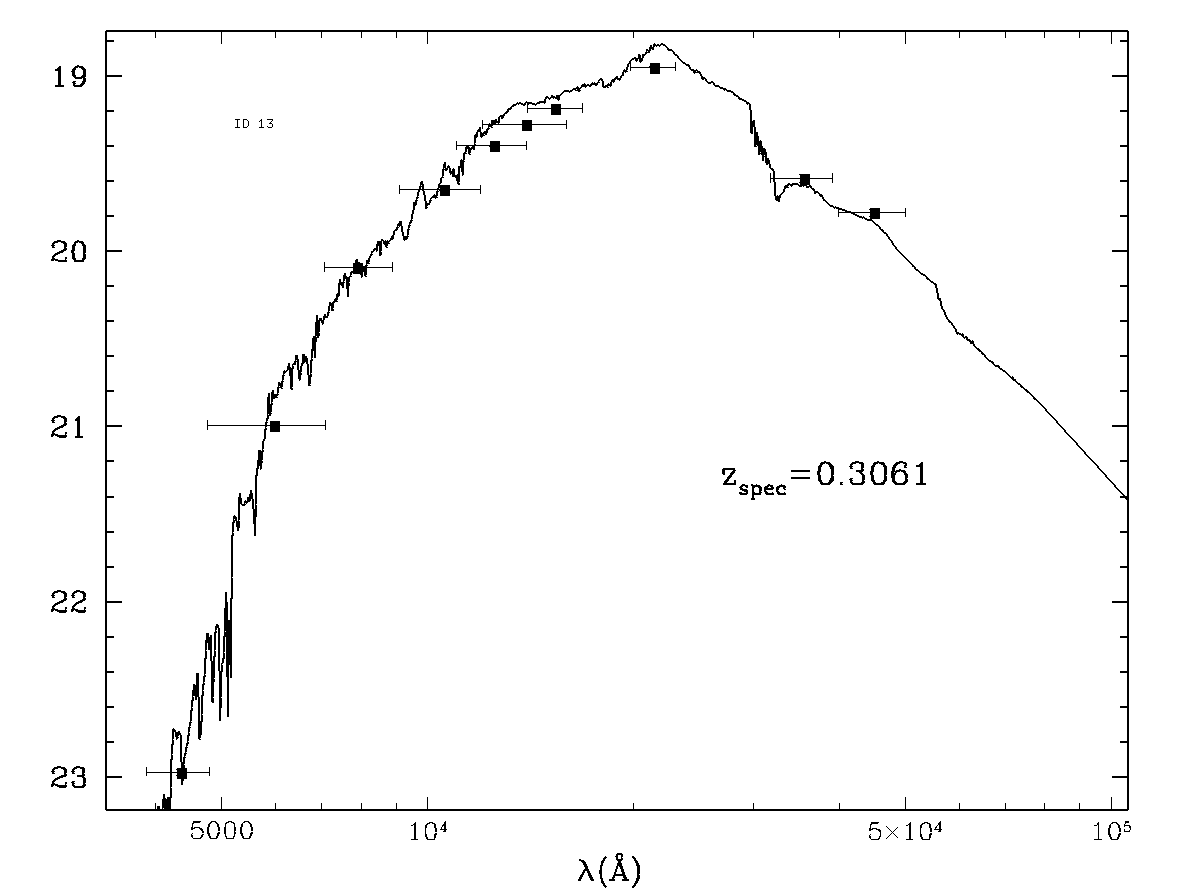}
   \includegraphics[width=7cm]{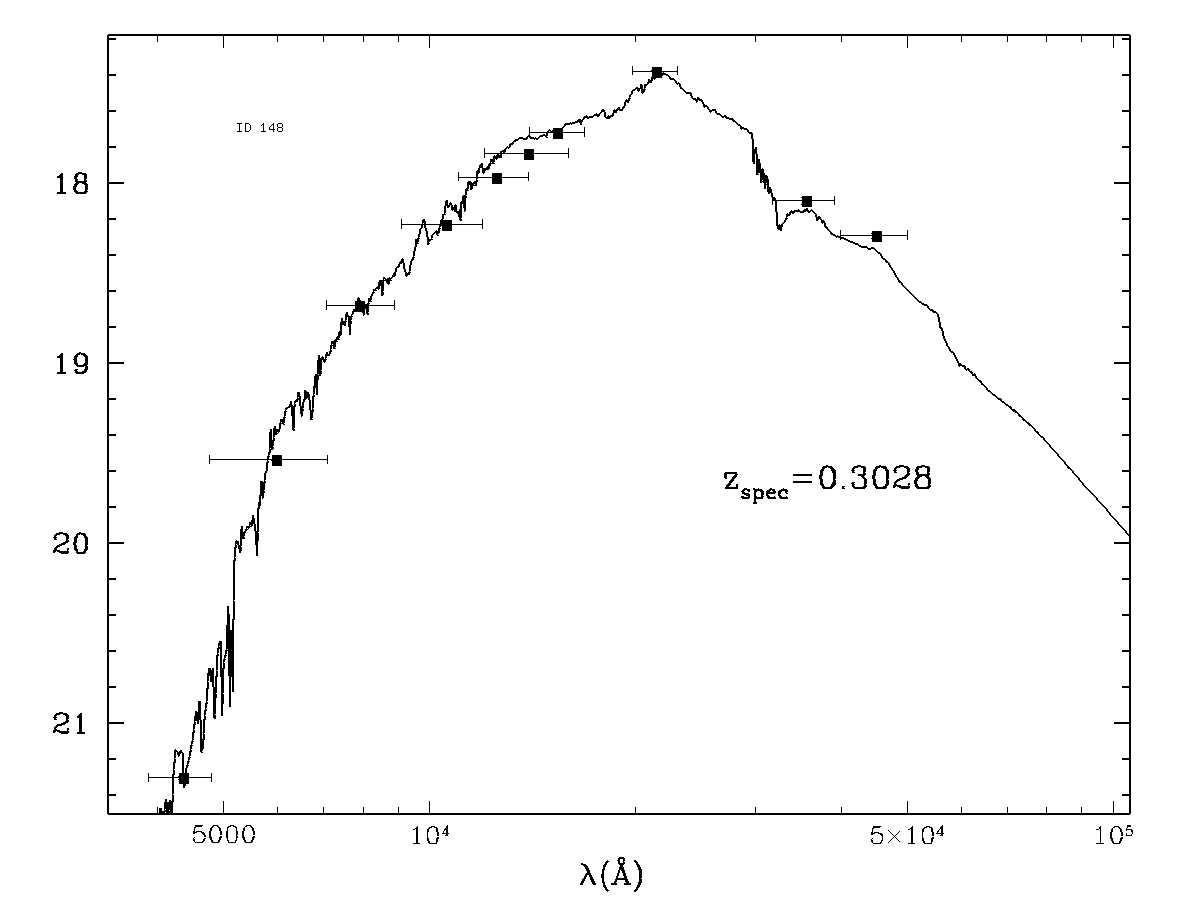}
   \includegraphics[width=7cm]{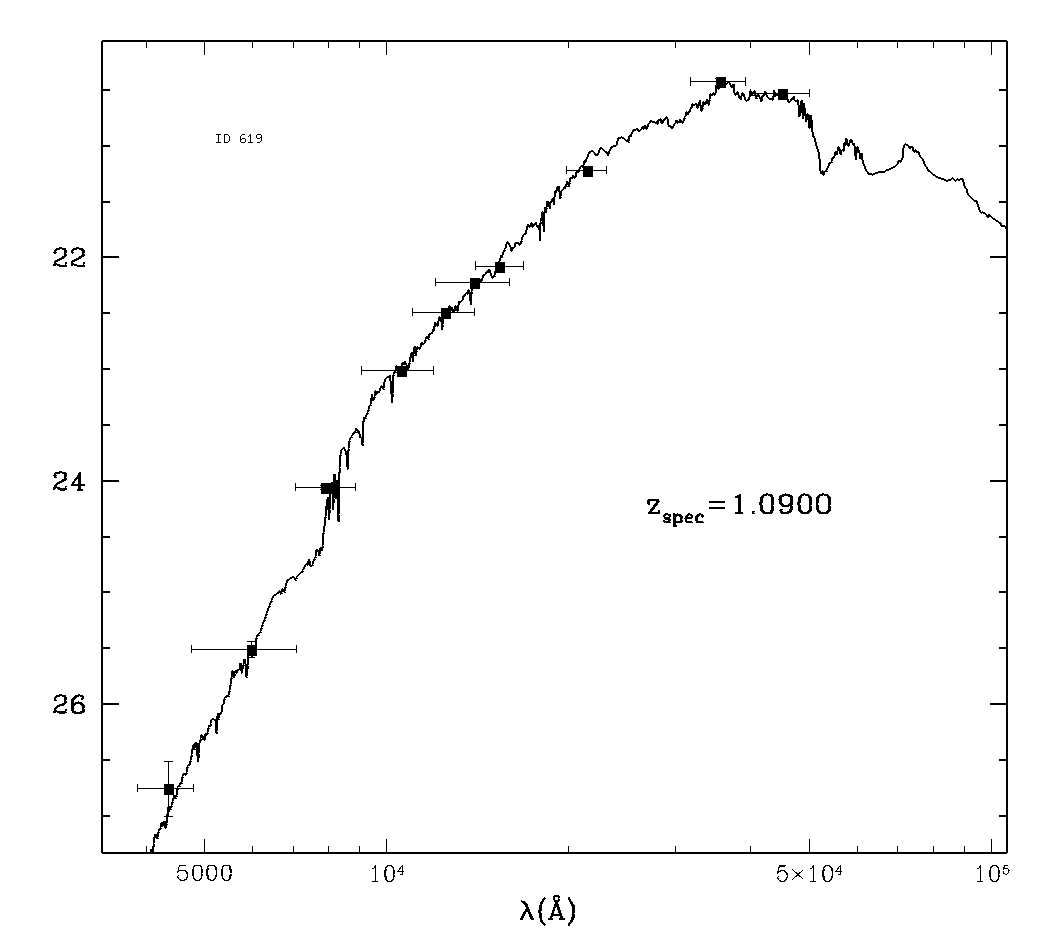}
    \includegraphics[width=7cm]{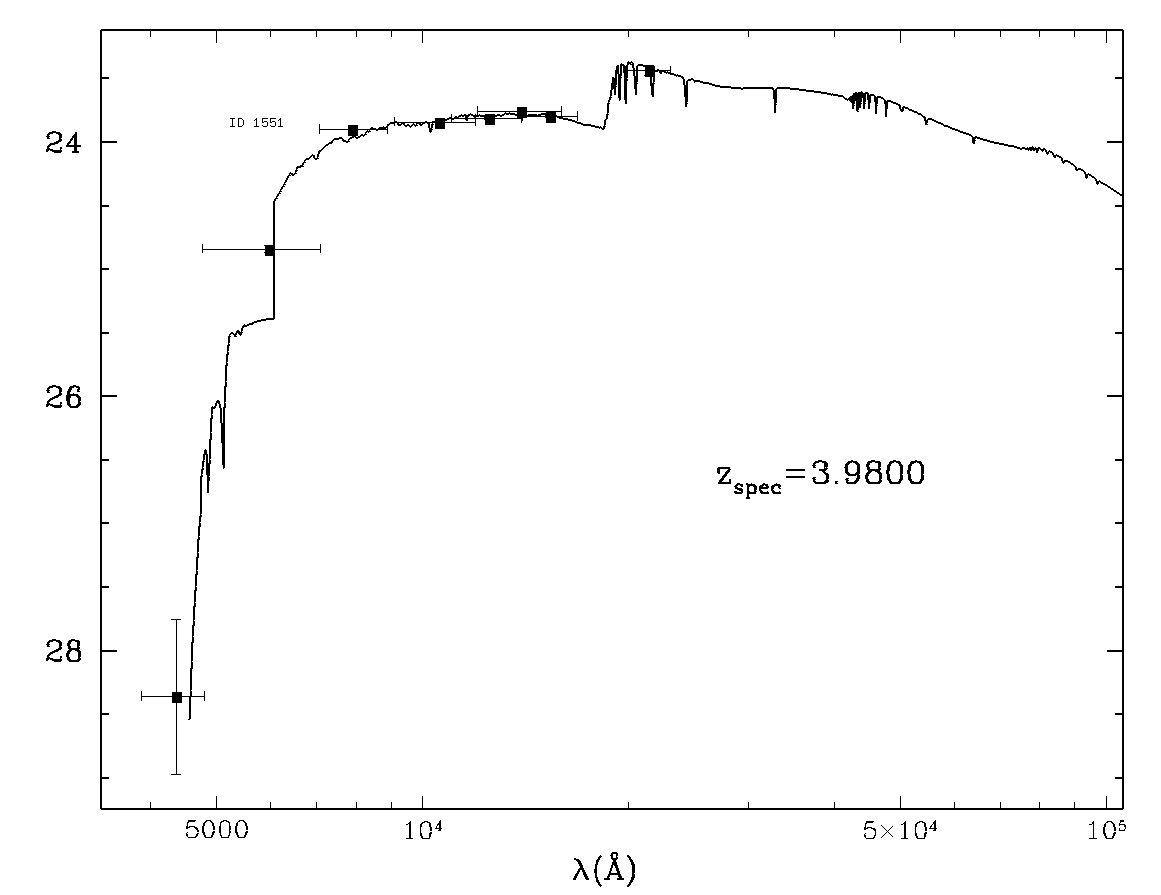}
   \centering
   \caption{Examples of SED fitting with photometry from this work, for four objects with known confirmed spectroscopic redshift in the A2744 cluster field.}  \label{speczSEDs}
\end{figure*}

  \begin{figure*}
   \centering
   \includegraphics[width=12cm]{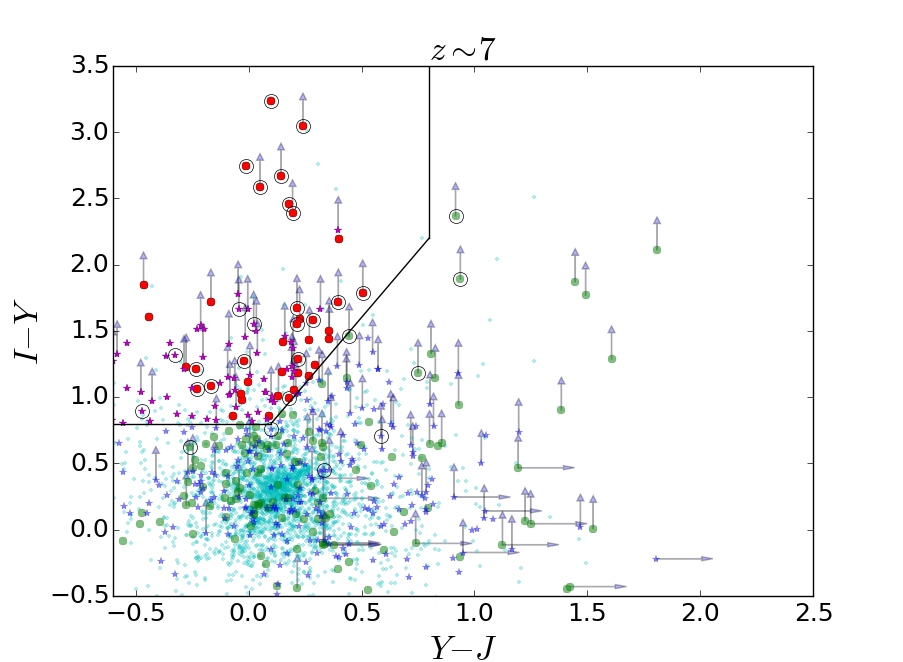}
   \includegraphics[width=12cm]{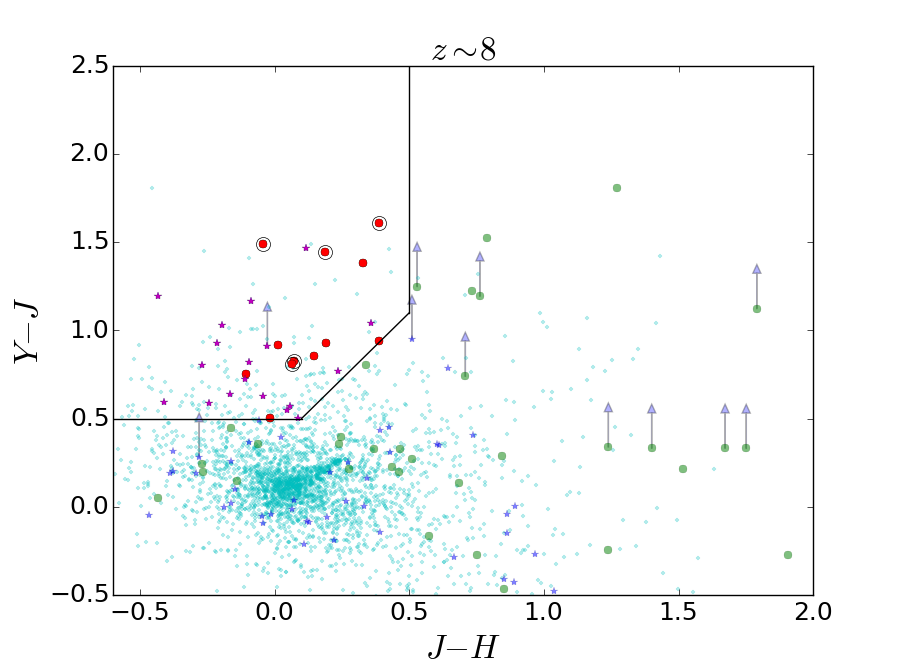}
   \centering
   \caption{Color selection of high-redshift sources. Left panel: $IYJ$ diagram for $z\sim7$ candidates (cyan points: sources with 1$\sigma$ detection in $B_{435}$ or $V_{606}$, which we exclude from the selection; green dots: $H$-detected sources, undetected in $B_{435}$ and $V_{606}$; blue stars: $IR$-detected sources, undetected in $B_{435}$ and $V_{606}$; red dots: $H$-detected $z\sim7$ candidates; magenta stars: $IR$-detected $z\sim7$ candidates; empty black circles: \citet{Atek2015} $z\sim7$ candidates; arrows represent upper limits). Right panel: $YJH$ diagram for $z\sim8$ candidates (symbols have the same meaning of the left panel, except that the cyan points represent sources that we exclude from the selection because they are detected in $B_{435}$, in $V_{606}$, or in $I_{814}$).} \label{colorselz}
\end{figure*}

Fig. \ref{rawcounts} presents the differential number counts for the four considered fields, as a function of the measured $H$160 magnitude. Fluxes are corrected for Galactic extinction derived from \citet{Schlegel1998} dust emission maps\footnote{We used the on-line calculator \textit{https://ned.ipac.caltech.edu/forms/calculator.html} to compute the extinction in the passbands of interest, at the central positions of the four fields.}. We also include the counts from the IR-detected catalogue. The four fields are in good agreement with one another, and show an overall overabundance with respect to the number counts on a re-scaled area of the CANDELS fields (GOODS-South and UDS are plotted in the Figures, for comparison). This effect is perfectly consistent with the expected cluster overdensity and magnification effects (which are present also in the parallel fields, although with relatively lower strength). Indeed, it can be shown that taking into account these effects (removing objects at the clusters redshift and de-magnifying the sources consistently with a cluster mass model) the number counts turn out to be in very good agreement with the CANDELS counts. This is discussed in full details in C16.

In Fig. \ref{colors} three color-color plots from the A2744 cluster field ($H$-detected) are shown; the other three fields have similar behaviour. The FF sources are distributed in overall good agreement with the CANDELS objects, with the noticeable exception of a sequence of color-clustered objects in each of the plots. From the photometric redshifts analysis performed in C16, these sources turn out to be mainly cluster members.

Fig. \ref{modelsSEDs} presents the fitted Spectral Energy distribution (SED) of three models of bright A2744 cluster objects (the SED fittings are obtained using the Fontana's photo-$z$ code developed at OAR, see C16 for all the details). The fluxes are obtained with \textsc{Galfit} for the HST bands, and with \textsc{t-phot} for the $Ks$ and IRAC bands. Fig. \ref{speczSEDs}, on the other hand, shows the fitted SEDs of four objects having known confirmed spectroscopic redshift \citep{Owers2011, Johnson2014}. In this case the fluxes are obtained with the \textsc{SExtractor} PSF-matched photmetric measurements for the HST bands, and with \textsc{t-phot} for the $Ks$ and IRAC bands. All the fits are quite accurate, showing that the applied recipes are efficient in retrieving reasonable fluxes over the whole range of considered wavelengths.

Finally, Fig. \ref{colorselz} shows a tentative color selection of high-redshift candidates based on our photometry. Following \citet{Atek2015}, the left panel shows the $IYJ$ diagram where we search for $z \sim 7$ sources by means of the following selection (we only consider objects that are not flagged as a potentially spurious detection resulting from the foreground light subtraction procedure, see Section \ref{detstr}):

\begin{itemize}
\item no detection in $B_{435}$ and $V_{606}$ 
\item $(I_{814}-Y_{105})>0.8$
\item $(I_{814}-Y_{105})>0.6+2\times(Y_{105}-J_{125})$
\item $(Y_{105}-J_{125})<0.8$
\end{itemize}

The right panel shows the $YJH$ diagram where we search for $z \sim 8$ objects by means of the following selection:
\begin{itemize}
\item no ``residual flag'' (see Section \ref{detstr})
\item no detection in $B_{435}$, $V_{606}$ and $I_{814}$ 
\item $(Y_{105}-J_{125})>0.5$
\item $(Y_{105}-J_{125})>0.4+1.6\times(J_{125}-H_{140})$
\item $(J_{125}-H_{140})<0.5$
\end{itemize}

We compare our results with those from the similar selection by \citet{Atek2015} (their high-redshift candidates are marked with large open circles in the two color-color diagrams, after a process of spatial cross-correlation between the two catalogs). We find an overall good agreement between the two sample selections, in both redshift intervals, with the noticeable exceptions of 8 sources (out of 29) in the $z\sim7$ diagram which fall shortly outside of the selection region when using our photometry. In C16 a comparison between the photometric redshifts derived from our photometry and other redshift estimates found in the literature is presented and discussed, finding a very good agreement for high-$z$ sources including those in the \citet{Atek2015} sample.  
On the other hand, we identify a number of new candidates, many of which (particularly in the $z\sim8$ selection) are $IR$-detected sources. In total, with this method we find 107 $z\sim7$ candidates (85 of which are $IR$-detected) and 31 $z\sim8$ candidates (28 of which are $IR$-detected). At a visual inspection, all these objects appear as reasonable high-redshift candidates. A thorough analysis of the topic is beyond the aim of this paper, and we leave it for future studies; however, this basic sanity check helps strengthening the reliability of our photometry.

\section{Summary and conclusions} \label{conclusions}

We have presented the complete multiwavelength photometric catalogues of the first two publicly released FF datasets, Abell2744 and MACSJ0416, along with the methodology we adopted to obtain them. The catalogs cover four fields (namely, the two clusters fields and the two corresponding parallel fields). In each catalogue, we list the total fluxes of the \textit{H}-band and IR-stack detected sources, in 10 passbands: HST ACS \textit{B}435, \textit{V}606, \textit{I}814; HST WFC3 \textit{Y}105, \textit{J}125, \textit{JH}140 and \textit{H}160; VLT Hawk-I \textit{Ks} 2.146 $\mu$m (ground-based); and Spitzer IRAC 3.6 and 4.5 $\mu$m. To detect the faint objects outshined by the bright foreground sources and the ICL in the two cluster fields, we develop a procedure to remove the light coming from these sources in the detection image (\textit{H}-band or IR-stack), fitting their light profiles with analytical models by means of the two public codes \textsc{Galapagos} and \textsc{Galfit}, applying a median filter to the processed images, and using \textsc{SExtractor} with a HOT+COLD approach. The parallel fields are processed with a more straightforward approach, directly running \textsc{SExtractor} on the {H}-band image.

The photometry in the HST bands is obtained using the \textsc{SExtractor} dual-mode option on PSF-matched images (convolved down to the \textit{H} band resolution). All HST cluster images are also processed as the detection image, removing the light from bright foreground sources. \textit{K} and IRAC fluxes are measured with \textsc{t-phot}, including an ad-hoc option to subtract local background light. In the case of the cluster fields, real sources and analytical models of bright objects are fitted simultaneously.

The catalogues also report the uncertainties on the flux measurements, as computed by means of the different techniques adopted to measure the fluxes, plus some additional diagnostic information (flags).

The first scientific applications of the catalogues, including photometric redshifts, magnification and physical properties, are presented in the companion paper by \citet{Castellano2016}.

The catalogues, together with the final processed images for all HST bands (as well as some diagnostic data and images) are publicly available and can be downloaded from the \textsc{Astrodeep} website at \texttt{http://www.astrodeep.eu/frontier-fields/}, and from a dedicated CDS webpage (\texttt{http://astrodeep.u-strasbg.fr/ff/index.html}).

\begin{acknowledgements}
The research leading to these results has received funding from
the European Union Seventh Framework Programme (FP7/2007-2013) 
under grant agreement n. 312725.
JSD acknowledges the support of the European Research Council via the award of an Advanced Grant.
FB acknowledges support by FCT via the postdoctoral fellowship SFRH/BPD/103958/2014 and also the funding from the programme UID/FIS/04434/2013.
RJM acknowledges the support of the European Research Council
via the award of a Consolidator Grant (PI McLure).
\end{acknowledgements}

\bibliographystyle{aa}

\begin{appendix}

\section{SExtractor parameters for detection} \label{sexparams}

Table \ref{detpar} lists the parameters adopted in the \textsc{SExtractor} detection runs described in Section \ref{detstr}.

\begin{table}[h!]
\begin{center}
\begin{tabular}{ | l | l | l |}
\hline
\textbf{Parameter} & \texttt{COLD} & \texttt{HOT} \\ \hline
\texttt{DETECT\_MINAREA} & 10.0 & 6.0 \\ \hline
\texttt{DETECT\_THRESH} & 0.7 & 0.82 \\ \hline
\texttt{ANALYSIS\_THRESH} & 0.7 & 0.82 \\ \hline
\texttt{DEBLEND\_NTHRESH} & 64 & 64 \\ \hline
\texttt{DEBLEND\_MINCOUNT} & 0.0001 & 0.0001 \\ \hline
\texttt{BACK\_SIZE} & 128 & 32 \\ \hline
\texttt{BACK\_FILTERSIZE} & 1 & 3 \\ \hline
\texttt{BACKPHOTO\_TYPE} & \texttt{local} & \texttt{local} \\ \hline
\texttt{BACKPHOTO\_THICK} & 48.0 & 48.0 \\ \hline
\texttt{MEMORY\_OBJSTACK} & 400 & 400 \\ \hline
\texttt{MEMORY\_PIXSTACK} & 4000000 & 4000000 \\ \hline
\texttt{MEMORY\_BUFSIZE} & 500 & 500 \\ \hline
\end{tabular}
\caption{\textsc{SExtractor} \texttt{COLD} and \texttt{HOT} mode parameter sets. A gaussian filtering (FWHM = 4.0 pixels) was applied for the detection runs.} \label{detpar}
\end{center}
\end{table}

\section{Catalogues formats} \label{catformat}

All the catalogues and derived quantities described in this paper are publicly released and can be downloaded from the \textsc{Astrodeep} website at
\texttt{http://www.astrodeep.eu/frontier-fields/}. The catalogues and images can be browsed from a dedicated interface at \texttt{http://astrodeep.u-strasbg.fr/ff/index.html}.

In all the catalogues the IDs are organized as follows:
\begin{itemize}
\item \textit{H}-detected objects have IDs starting from 1;
\item IR-detected objects have IDs starting from 20000;
\item the bright cluster objects, modeled and subtracted from the HST images, have IDs starting with 100000.
\end{itemize}

The formats of the released catalogs are as follows:
\begin{itemize}
\item \textbf{Catalogue A - magnitudes}: in the first catalogue we list IDs, position, AB magnitudes and relative uncertainties, in the ten considered bands. The format is therefore\\ \texttt{ID RA DEC X Y MAG\_B435 MAG\_V606 MAG\_I814 MAG\_Y105 MAG\_J125 MAG\_JH140 MAG\_H160 MAG\_Ks MAG\_IRAC1 MAG\_IRAC2 MAGERR\_B435 MAGERR\_V606 MAGERR\_I814 MAGERR\_Y105 MAGERR\_J125 MAGERR\_JH140 MAGERR\_H160 MAGERR\_Ks MAGERR\_IRAC1 MAGERR\_IRAC2}.
\item \textbf{Catalogue B - fluxes}: a second catalogue contains IDs, fluxes and uncertainties of the fluxes ($\mu Jy$). The format is\\ \texttt{ID FLUX\_B435 FLUX\_V606 FLUX\_I814 FLUX\_Y105 FLUX\_J125 FLUX\_JH140 FLUX\_H160 FLUX\_Ks FLUX\_IRAC1 FLUX\_IRAC2 FLUXERR\_B435 FLUXERR\_V606 FLUXERR\_I814 FLUXERR\_Y105 FLUXERR\_J125 FLUXERR\_JH140 FLUXERR\_H160 FLUXERR\_Ks FLUXERR\_IRAC1 FLUXERR\_IRAC2}.
\item \textbf{Catalogue C - diagnostics}: a third catalogue contains useful diagnostic data. It lists IDs, position in $H$160 image pixel reference, segmentation limits, \textsc{SExtractor} \texttt{CLASS\_STAR} parameter, the flag we applied to residual features after processing the detection image (see Section \ref{detstr}), a "visual inspection flag" to select spurious detections, the flag given by \textsc{t-phot} to identify saturated and blended priors, and the information on the covariance of the sources (ratio of maximum covariance to the variance) for the $Ks$ and IRAC bands. The format is \\ \texttt{ID X Y XMIN YMIN XMAX YMAX CLASS\_STAR SEXFLAG RESFLAG VISFLAG \\ TPHOTFLAG\_Ks COVMAX\_Ks TPHOTFLAG\_IRAC1 COVMAX\_IRAC1 \\ TPHOTFLAG\_IRAC2 COVMAX\_IRAC2]}.
\end{itemize}
\end{appendix}

\end{document}